\documentclass[12pt]{article}
\usepackage{amssymb}
\usepackage{epsfig}
\usepackage{sint,cite}
\usepackage{bbold}

\newcommand{\nn}{\nonumber}

\newcommand\be{\begin{equation}}
\newcommand\ee{\end{equation}}
\newcommand\bes{\begin{eqnarray}}
\newcommand\ees{\end{eqnarray}}
\newcommand{\bea}{\begin{array}}
\newcommand{\ea}{\end{array}}
\newcommand{\eq}[1]{eq.~(\ref{#1})}
\newcommand{\fig}[1]{Fig.~\ref{#1}}
\newcommand{\tab}[1]{Table~\ref{#1}}

\newcommand{\ev}[1]{\langle #1 \rangle}
\def\mdm{M^{\dagger}M}
\def\mdme{(\mdm)_e}
\def\mdmo{(\mdm)_o}
\newcommand\norm[1]{\|#1\|}
\def\rloc{r_{\rm loc}}
\def\rlocm{r_{\rm loc}^{\rm max}}
\def\mG{m_{\rm G}}
\def\nmin{n_{\rm min}}
\def\tr{{\rm tr}}
\def\bpsi{\bar{\psi}}
\def\seff{S_{\rm eff}}
\def\bchi{\bar{\chi}}
\newcommand\Enorm[1]{\norm{#1}_{\rm E}}
\def\phid{\phi^{\dagger}}

\def\ma[#1,#2,#3,#4]  {{\left( \matrix{ #1  & #2 \cr
                                        #3  & #4 \cr } \right)}}

\begin{document}

\thispagestyle{empty}
\title{{\normalsize\vskip -50pt
\mbox{} \hfill HU-EP-04/16 \\
\mbox{} \hfill DESY 04-048  }
\vskip 25pt
%On the problem of formulating a local theory
%when reducing the number of staggered fermion tastes
Locality with staggered fermions
}

\author{
B. Bunk $^{a}$, M. Della Morte $^{a}$, K. Jansen $^{b}$ and
F. Knechtli $^{a}$\\[1cm]
$^{a}$ Institut f\"ur Physik, Humboldt Universit\"at,\\
Newtonstr. 15, 12489 Berlin, Germany\\[0.5cm]
$^{b}$ NIC/DESY-Zeuthen,\\ Platanenallee 6, 15738 Zeuthen, Germany\\[1cm]
}

\maketitle

\begin{abstract}
 We address the locality problem arising in simulations, which take the
 square root of the staggered fermion determinant as a Boltzmann weight
 to reduce the number of dynamical quark tastes.
 A definition of such a theory necessitates an underlying
 {\em local} fermion operator
 with the same determinant and the corresponding Green's functions
 to establish causality and unitarity.
 We illustrate this point by studying analytically and numerically the
 square root of the staggered fermion operator. Although it has the
correct
 weight, this operator is non-local in the continuum limit.
 Our work serves as a warning that fundamental properties of field
theories
 might be violated when employing blindly the square root trick.
 The question, whether a local operator reproducing the square root of the
 staggered fermion determinant exists, is left open.
\end{abstract}

\newpage

\section{Introduction}

Traditionally two formulations of lattice QCD, Wilson \cite{Wilson:1975id} and
staggered or Kogut-Susskind fermions \cite{Kogut:1975ag,Susskind:1977jm},
were used extensively in practical simulations.
Wilson type fermions suffer from the breaking of chirality.
Staggered fermions
rely on the appealing idea to get rid of explicit spin degrees of freedom. 
They arise naturally constructing the square root of the lattice Laplace operator
\cite{Rossi:1984kr}. Massless staggered fermions have a continuous
taste non-singlet U(1) axial symmetry which is dynamically
broken yielding one Goldstone pion
\cite{Kluberg-Stern:1981wz,Kawamoto:1981hw,Kluberg-Stern:1983dg,Golterman:1984cy,Kilcup:1987dg}.
This symmetry prevents from additive mass renormalization and together with
the reduced number of degrees of freedom per lattice site, these properties
make staggered fermions well suitable for numerical simulations. 
Their ``unwanted'' properties are the doubling, in the continuum limit each flavor
comes in four ``tastes'',
the breaking of the taste symmetry at finite lattice spacing and
the not straightforward construction of operators.

In addition to Wilson and staggered fermions (and their improved versions)
many new discretizations of fermions on the
lattice have been proposed, most notably overlap fermions \cite{Neuberger:1998fp},
see Ref. \cite{Jansen:2003nt} for a recent review.

With staggered fermions many computations within the quenched approximation
have been performed \cite{Toussaint:2001zc,Bernard:2001av}.
An interesting and important universality check
is represented by the APE plot shown in Ref. \cite{Gattringer:2003qx}
where quenched data from different fermion discretizations are compared
to the continuum extrapolation of the CP-PACS collaboration data \cite{Aoki:2002fd}.

Since the early times of lattice computations staggered fermions have been simulated
dynamically \cite{Gottlieb:1987mq}. The doubling in four degenerate tastes in
the continuum limit can be exploited since the tastes behave like quarks as far as the
QCD interaction is concerned. But to describe real world QCD one needs a formulation
for one or two tastes. An early attempt to describe two tastes by reducing the
degrees of freedom, called the reduced staggered fermion formalism
\cite{Sharatchandra:1981si}, although it satisfies unitarity and positivity,
leads to a complex determinant \cite{vandenDoel:1983mf} and it is therefore not
suitable for numerical simulations.
Another way of reducing the number of tastes
is the so called square (or fourth) root trick \cite{Marinari:1981qf}. It
amounts of taking the square (or fourth) root of the staggered fermion determinant,
motivated by its factorization in the continuum limit. While in several places
warnings about its potential danger have been expressed
\cite{Marinari:1981qf,Davies:2003ik},
the central problem that this is not an {\em ab initio} formulation of lattice QCD
\cite{Jansen:2003nt,DeGrand:2003xu,Neuberger:2004ft}
has not been addressed in the literature.
Arguments based on partially quenched chiral perturbation theory
\cite{Bernard:1994sv}
support the square root trick but this should only be considered as a motivation
for a first principle study. 

From the numerical simulation side there are no major obstacles to implement
the square root trick. But as more and more results based on such simulations
are being published and ambitious high precision tests of the standard model
announced \cite{Davies:2003ik}, the theoretical issues postponed so far have
to be addressed. Here we only try to state the problem concerning locality,
following a benchmark investigation for the overlap case in Ref.
\cite{Hernandez:1998et}.

At a glance the problem discussed in this work can be introduced by the following
simple arguments.
\begin{enumerate}
\item In the QCD partition function on the lattice only local
operators appear
\bes
 \mathcal{Z} & = & \int_{U,\bpsi,\psi}
                   \exp\left\{-S_g(U)+a^4\sum_x\bpsi(x)D\psi(x)\right\}
                   \,,\label{level1}
\ees
where $S_g(U)$ is the gauge action in terms of gauge links
$U$ and $D$ is the, assumed to be local, Dirac operator acting on fermion fields
$\bpsi,~\psi$.
\item The integration over the fermion fields
$\bpsi,~\psi$ generates an effective action
\bes
 \mathcal{Z} & = & \int_U\det(aD)\exp\left\{-S_g(U)\right\} =
                   \int_U\exp\left\{-S_g(U)+\tr\ln(aD)\right\}
                   ,\label{level2}
\ees
which is {\em non-local} in the gauge link variables.
\end{enumerate}
The non-locality of the effective action does not mean a non-locality of the theory
if the latter possesses an underlying local formulation in terms of
{\em fundamental} degrees of freedom.

If the starting point of a lattice theory is \eq{level2}, e.g. with the
effective action
\bes
 \seff & = & -S_g(U)+\frac{1}{2}\tr\ln(aM) \label{sqrtaction}
\ees
corresponding to the Boltzmann weight $\sqrt{\det(aM)}$, then
an equivalent formulation like in \eq{level1} in terms of a local operator $D$
is needed in order to establish causality and discuss renormalizability and
universality.

The present work is organized as follows.
In Section 2 we formulate precisely the locality problem
related to the square root trick. We present our ``candidate'' for a local
operator $D$ namely the most naive choice, which is to take the square root
of the staggered operator. In Section 3 we present our analytical results,
in particular in the free field theory where the square root operator is proven
to be non-local. Numerical investigations are still of interest to establish the
actual ``non-localization'' range in the interacting theory. Section 4 presents
the simulation results in the quenched approximation. We discuss these results
in Section 5. There are three appendices devoted to the derivation of our
analytical results and to the study of finite size effects in our simulations.

\section{Formulation of the problem}

We consider an Euclidean hypercubic lattice with lattice spacing $a$ and coordinates
$x_\mu=n_\mu a$ with $n_\mu=0,1,\ldots,L/a-1$. The number of sites in each
direction is $L/a$ and it is assumed to be even. The directions are denoted by
$\mu=0,1,2,3$ and $a_\mu$ is a displacement vector by one lattice spacing
along the $\mu$ direction.
The action for Kogut-Susskind staggered fermions is
\bes
 S & = & S_g(U)-a^4\sum_x\bchi(x)M\chi(x) \label{stagga} \\
 M\chi(x) & = & m\chi(x) + \sum_\mu\frac{1}{2a}\eta(x,\mu)\left[
                U(x,\mu)\chi(x+a_\mu) - 
                U^\dagger(x-a_\mu,\mu)\chi(x-a_\mu)
                \right] \,, \nonumber
\ees
where $\eta(x,\mu) = (-1)^{\sum_{\nu<\mu}n_\nu}$ are the staggered phases.
The fermion field $\chi$ has only SU(3) color indices. Periodic boundary
conditions are used for the gauge field and (anti-)periodic for the fermion field.
In the continuum limit $M$ exhibits spectrum doubling since it describes four
degenerate ``tastes'', which can be interpreted as four flavors of quarks
\cite{Sharatchandra:1981si,Kluberg-Stern:1983dg,Golterman:1984cy}. At finite
lattice spacing there are O($a^2$) taste-changing interactions even in the
free theory. The main advantage of staggered fermions is that the action 
\eq{stagga} is invariant for $m=0$ under the continuous taste non-singlet
U(1) axial transformation
\bes
 \chi(x)\to{\rm e}^{i\beta\epsilon(x)}\chi(x)\,,\qquad
 \bchi(x)\to{\rm e}^{i\beta\epsilon(x)}\bchi(x)\,,
\ees
where
\bes
 \epsilon(x) & = & (-1)^{\sum_\mu n_\mu} \,, \label{epseo}
\ees
from which it follows that there is no additive mass renormalization
\cite{Sharatchandra:1981si,Golterman:1984cy}.

The question, whether there exists a formulation for two degenerate tastes of
staggered fermions has been considered in Ref. \cite{Sharatchandra:1981si}. 
In this so called reduced staggered fermion formalism
the fermion field $\chi$ lives on odd sites only and
the anti-fermion field $\bchi$ on even sites only
(``even'' and ``odd'' being defined by the sign of $\epsilon(x)$ \eq{epseo}).
Although this theory satisfies unitarity and positivity it leads
to a complex fermion determinant \cite{vandenDoel:1983mf}
and is therefore not suitable for numerical Monte Carlo studies.
Also this reduced formalism does
not have any continuous U(1) axial symmetry
\cite{vandenDoel:1983mf,Golterman:1984cy}.

The authors of Ref. \cite{Marinari:1981qf} proposed to take the square (or fourth)
root of the fermion determinant to reduce the number of tastes
from four to two (or one). Based on the
factorization in the naive continuum limit
\bes
 \det(aM) & \stackrel{a\to0}{\longrightarrow} & \det(a\Omega)^4
\ees
in terms of a one-flavor fermion operator $\Omega$, taking the square
root amounts to quenching two out of four tastes and this idea seems to
work out in partially quenched chiral perturbation theory \cite{Bernard:1994sv}.

The square (and fourth) root trick is employed in large scale simulations by the
MILC \cite{Gottlieb:2003bt} and JLQCD \cite{Aoki:2002xi} collaborations.
From the algorithmic side the
square root of the determinant can be dealt with the R algorithm
\cite{Gottlieb:1987mq}, with the PHMC algorithm \cite{Frezzotti:1997ym,Aoki:2002xi}
or with the RHMC algorithm \cite{Clark:2003na}.
In the computation of quark propagators the original four taste staggered operator
is used.
Correlation functions are built using sources that project onto
the desired valence taste components \cite{DeGrand:2003xu}.
This is justified in partially quenched chiral perturbation theory
with some complication for taste-singlet operators \cite{Bernard:1994sv}.
However the danger with such an approach are possible unitarity violations
through a mismatch of sea and valence quark lines.
This fundamental question, which
deserves more investigation, is not addressed in the present work.

In order to establish causality and universality for the theory
implicitly defined by taking the square root of the staggered fermion determinant,
a {\em local} definition of the theory from first principles like in \eq{level1}
is needed. Explicitly the problem is to find a fermion operator $D$ such that
\bes
 \det(aD) = \sqrt{\det(aM)} & \mbox{and} &
 \norm{G(x,y)}\le C{\rm e}^{-\gamma\Enorm{x-y}/a}\,, \label{locality}
\ees
with $C$ and $\gamma>0$ independent of $U$ 
\cite{Niedermayer:1998bi,Hernandez:1998et}.
In \eq{locality} $\norm{G(x,y)}$ is the operator norm of the kernel
\bes
 aD\psi(x) & = & a^4\sum_yG(x,y)\psi(y) \label{kernel}
\ees
and $\Enorm{x-y}$ is the Euclidean norm.

A technical point is that
so far we have used $M$ as the operator describing four tastes.
In numerical simulations the fermion determinant is represented by
pseudofermions and for this the Hermitean positive definite operator $\mdm$
is needed, which describes eight tastes. The number of tastes can be reduced
to the original four by noting that
$\mdm$ decouples the even and the odd sublattices
and that \cite{Martin:1985yn}
\bes
 \det(aM) & = & \det(a^2\mdme) =\det(a^2\mdmo) \,. \label{detmo}
\ees
In \eq{detmo} the subscripts for $\mdme$ and $\mdmo$ 
refer to the operator $\mdm$ acting on fields living only on the {\em even} or
{\em odd} sublattices respectively. In the following
to describe four tastes of staggered fermions the operator
$\mdme$ is used.

In this work we investigate the locality properties of the so far only known
candidate for an operator $D$ to satisfy \eq{locality}, namely
\bes
 D & = & \sqrt{(\mdm)_e} \,, \label{ourD}
\ees
where the operator square root is obtained by a Chebyshev polynomial approximation
\cite{Numerical-Recipes}. It approximates the unique \cite{Matrix-Analysis}
Hermitean positive definite square root of $\mdme$.

\section{Analytical results}

\subsection{Bound}

To derive a bound on the locality of the operator $D$ defined in \eq{ourD}
we need to know
the spectral bounds of $\mdm$ \cite{Kalkreuter:1995ax}. In the free theory
\bes
  a^2\mdm{\rm e}^{ipx} & = & [(am)^2 + \sum_\mu\sin^2(p_\mu a)]
                             {\rm e}^{ipx} \label{momrep}
\ees
which implies for the spectrum 
${\rm spec}(a^2\mdm)\subset[(am)^2,4+(am)^2]$.
In the interacting case the upper bound is lifted\footnote{
The upper bound is equivalent to the one for naive fermions.}
\bes
  {\rm spec}(a^2\mdm) & \subset & [u,v]\,,\quad
  u=(am)^2\,,\quad v=16+(am)^2 \,. \label{specbound}
\ees
The operator $\mdme$ has the same spectral bounds \cite{Martin:1985yn}.

In order to handle analytically the operator $D$
we would like to use the series of polynomials $S_n(z)$ of degree $n$ 
defined by the generating function
\bes
 \sqrt{1+t^2-2tz} & = & \sum_{n=0}^{\infty}S_n(z)t^n \,. \label{genfun}
\ees
Properties of this series (where $z$ is a number)
are discussed in the Appendix \ref{polyexp}.
Following Ref. \cite{Hernandez:1998et} we define the operator (with
obvious insertions of unit matrices)
\bes
 z & = & \frac{v+u-2a^2\mdme}{v-u} \,, \label{zdef}
\ees
which maps the spectrum $[u,v]$ of $a^2\mdme$ into $[-1,1]$. We then set
\bes
 t={\rm e}^{-\theta} & \mbox{with} &
 \cosh\theta = \frac{v+u}{v-u} \,,\quad \theta>0 \,. \label{tandtheta}
\ees
For this choice of $t$ we have
\bes
 1+t^2-2tz & = & \frac{4{\rm e}^{-\theta}}{v-u}a^2\mdme
\ees
and we can use \eq{genfun} to provide a series expansion for
the operator $D$.

The kernel $G(x,y)$ \eq{kernel} of the square root operator
can be expressed in terms of the kernels $G_n(x,y)$ for
the polynomial operators $S_n(z)$. From \eq{genfun} and
\eq{tandtheta} we get
\bes
 G(x,y) & = & \frac{1}{2}\sqrt{v-u}{\rm e}^{\theta/2}
 \sum_{n=0}^{\infty}G_n(x,y){\rm e}^{-n\theta}
\ees
Since $z$ as it is defined in \eq{zdef} connects two neighboring even sites
on the lattice, we have
\bes
 G_n(x,y)=0 & \mbox{for} & n<\nmin=\frac{\norm{x-y}_1}{2a} \,,\nonumber
\ees
where $\norm{x-y}_1/a$ is the number of links between $x$ and $y$
({\em ``taxi driver distance''}). In \eq{Om-est2} a bound for the
remainder of the polynomial series in \eq{genfun} truncated after
$n=\nmin-1$ terms is given. Since the spectrum of $z$ \eq{zdef}
is contained in $[-1,1]$ and using the operator calculus, the bound
\eq{Om-est2} implies
\bes
 a^4\norm{G(x,y)} & \le & 
 \frac{\sqrt{v-u}}{(\nmin-1)\pi}{\rm e}^{\theta/2}
 {\rm e}^{-\norm{x-y}_1/(2a/\theta)} \,,\quad \nmin\ge2 \label{bound}
\ees
for the norm of the kernel $G(x,y)$ in SU(3) color space. Inserting
the values for the spectral bounds given in \eq{specbound} into the
definition of $\theta$ \eq{tandtheta} we obtain
\bes
 \theta & = & \frac{am}{2}+{\rm O}((am)^3) \,.
\ees
This means that $a^4\norm{G(x,y)}$ is bounded by an exponential
$\propto\exp(-r/\rloc)$ with $r=\norm{x-y}_1$ and
\bes
 \rloc & = & \frac{2a}{\theta}\simeq\frac{4}{m} \,. \label{rlocbound}
\ees
The localization range $\rloc$
stays {\em finite} in the continuum limit.
This is in contradiction to a local theory where we need $\rloc$ proportional to
the lattice spacing.

The result \eq{bound} represents an {\em upper} bound on the localization of
the operator $D$, the theory could in principle still be local in the continuum.
We devote
therefore our attention to the free theory where we can achieve an exact result.
Since in the free theory $\theta=am+{\rm O}((am)^3)$, the bound \eq{bound}
gives $\rloc^{({\rm free})}\simeq2/m$.

\subsection{Exact result in the free theory}

We consider an infinite lattice. The free operator $\mdm$ decouples on sixteen
sublattices of lattice spacing $2a$ where it acts like a Laplace operator.
The Fourier transformation of the square root of the diagonal operator in
momentum space \eq{momrep} yields for the kernel $G(x,y)$ \eq{kernel}
\bes
 G(x,y) & = &
 \int_{-\pi/(2a)}^{\pi/(2a)}\frac{d^4p}{\pi^4}
 \sqrt{(am)^2+\sum_\mu\sin^2(p_\mu a)}\,{\rm e}^{ip(x-y)} \,,
 \label{latfree}
\ees
where $(x_\mu-y_\mu)/a$ is even for all $\mu$, that is
$x$ and $y$ live on one of the sixteen sublattices of lattice spacing $2a$.
Clearly then \eq{latfree} applies also for $D$ in \eq{ourD}. 
This integral is solved in Appendix \ref{sqrtfree} and the result 
(for $y=0$) is given by replacing $a$ with $2a$ in \eq{eq-rootx}.

The continuum version of \eq{latfree} is also computed in Appendix \ref{sqrtfree},
the result is \eq{eq-rootcont}
\bes
 &&\int_{-\infty}^{\infty}\frac{d^4p}{(2\pi)^4}\sqrt{p^2+m^2}
 \,{\rm e}^{ipx} = \nonumber \\
 &&-\frac{1}{4\pi^2}\frac{m^2}{\Enorm{x}^3}\left(
 1 + \frac{3}{m\Enorm{x}} + \frac{3}{m^2\Enorm{x}^2} \right)
 {\rm e}^{-m\Enorm{x}} \,. \label{freethcont}
\ees
As is shown in Appendix \ref{sqrtfree} the continuum and lattice results
agree at large Euclidean distance
and this establishes that the operator $D$ in \eq{ourD}
is {\em non-local} in the free continuum limit. The localization range
is $\rloc=1/m$. By noting that $\Enorm{x}\le\norm{x}_1\le2\Enorm{x}$
we see that the bound \eq{bound} in the free theory is saturated.

It is very unlikely that the
introduction of gauge interactions changes qualitatively this result.
From the numerical point of view it is still interesting to investigate
what actually happens in the interacting theory. Because of confinement
the quark mass will be ``replaced'' in the bound \eq{rlocbound} by some
hadronic mass, which could be very large in a favorable case.
From a similar study in the overlap case \cite{Hernandez:1998et}
it is known that the analytical bound is only poorly saturated.

\section{Numerical results}

To test numerically the locality of the operator $D$ \eq{ourD} we follow
closely Ref. \cite{Hernandez:1998et}.
We consider hypercubic lattices with $L/a$ sites in each direction and periodic
boundary conditions. The operator $D$ acts on fields
living on the even sites only. We define the source field
\bes
 \xi_c(x) & = & \left\{\bea{ll} 1\quad & \mbox{if $x=y$ and $c=1$} \\
                                      0      & \mbox{otherwise}\,, \ea\right.
 \label{source}
\ees
where $y$ is the location of the source and $c$ runs over the color index of the
field. We take a point-like color source
because we are only interested in the decay properties of
\bes
 \psi(x) & = & aD\xi_c(x) \label{applyD}
\ees
described in terms of the function
\bes
 f(r) & = & 
 \max\big\{\norm{\psi(x)}\,\big|\,\norm{x-y}_1=r\big\} \,. \label{fr}
\ees
A different choice of the source in \eq{source}, e.g. spread in a $2^4$ hypercube
describing ``one'' taste\footnote{
At finite lattice spacing there are taste changing interactions.}
will not change the result, which concerns a mathematical property of the
operator that needs to be fulfilled.
In \eq{fr} $\norm{\psi(x)}$ is the SU(3) color norm and the ``taxi driver distance''
is defined as
\bes
 \norm{x-y}_1 & = & \sum_{\mu}\min\left\{ 
 |x_\mu-y_\mu| , L-|x_\mu-y_\mu| \right\} \,. 
 \label{dtaxi}
\ees
%In \eq{dtaxi} the coordinate differences are computed
%taking into account the periodicity
%of the lattice so as to minimize
%the distance. 
Hence $\norm{x-y}_1/a$ is
the number of links for the shortest path on the lattice between $x$ and $y$
using the periodicity.
The maximal value that $\norm{x-y}_1$ can take is $2L$.
The taxi driver distance is used here since it arises naturally in the
derivation of the analytic bound \eq{rlocbound}.
\begin{table}[t] 
   \centering
   \begin{tabular}{|ccccccc|}
    \hline
    $\beta$ & $am$ & $r_0\mG$ & $L/a$ & $\varepsilon$ & pol. degree & \#configs.
  \\[0.5ex] \hline\hline
6.0 & 0.01  & 1.29 \cite{Kim:1995tg}
                      & 16    & $2\times10^{-7}$ & 490  & 200 \\
    &       &         & 24,32 & $8\times10^{-9}$ & 735  & 600 \\
  \hline
6.2 & 0.007 & 1.34(4) \cite{Gupta:1991mr}
                      & 24    & $4\times10^{-9}$ & 1050 & 150 \\
    &       &         & 36    & $4\times10^{-9}$ & 1050 & 300 \\
  \hline
6.5 & 0.005 & 1.27(1) \cite{Kim:1999ur}
                      & 32    & $2\times10^{-9}$ & 1470 & 300 \\
    &       &         & 48    & $2\times10^{-9}$ & 1470 & 192 \\
  \hline 
 \end{tabular} 
 \caption{Simulation parameters. The last three columns tabulate
 the relative accuracy $\varepsilon$ \eq{accu},
 the degree of the polynomial and the number of configurations generated.
 \label{t_par}}
\end{table}
\begin{figure}[t]
 \begin{center}
     \includegraphics[width=12cm]{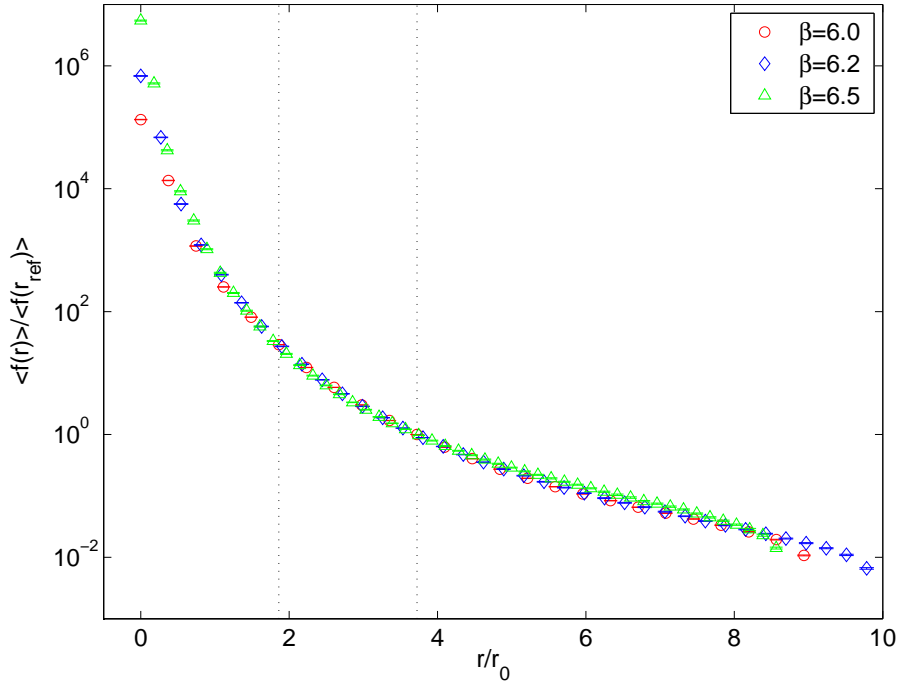}
 \end{center}
 \vspace{-0.5cm}
 \caption{Decay of $\ev{f(r)}$ defined in \eq{fr}.
 The curves are normalized to be 1 at the distance
 $r_{\rm ref}/r_0=3.72$ corresponding to the location of the rightmost
 vertical dotted line. \label{f_fr}}
\end{figure}

We compute the function $f(r)$ on configurations generated in the quenched
approximation and denote the average in the quenched ensemble by $\ev{f(r)}$.
We keep the location of the source fixed at $y=0$, the average over
gauge configurations projects onto the gauge invariant part of \eq{fr}.
\tab{t_par} summarizes the simulation parameters. We simulate
at three $\beta$ values in order to take the continuum limit.
The quark masses are obtained from the literature 
\cite{Gupta:1991mr,Kim:1995tg,Kim:1999ur}
and define a line of constant
physics where the mass $\mG$ of the Goldstone pion $\pi_{\rm G}$
is $r_0\mG=1.30(3)$ in units of the hadronic scale $r_0$
\cite{Sommer:1994ce,Guagnelli:1998ud}.
We have two approximately matched physical volumes $L\mG\approx4$ and
$L\mG\approx6$ at all $\beta$ values. At $\beta=6.0$ we also have a third larger
volume, which we used to study the finite volume effects.
Statistical errors of derived quantities were determined
by the method of Ref. \cite{Wolff:2003sm}.

We implemented the Chebyshev polynomial approximation using the
Clenshaw's recurrence formula \cite{Numerical-Recipes}. As can be seen from
\tab{t_par} for most of the computations a relative accuracy
$\varepsilon=10^{-8}-10^{-9}$ was required, which is defined as
\bes
 \varepsilon=\frac{\norm{(D^2-\mdme)R}}{\norm{\mdme R}} \label{accu}
\ees
where $R$ is a normalized Gaussian random vector $R$.
\begin{figure}[t]
 \begin{center}
     \includegraphics[width=12cm]{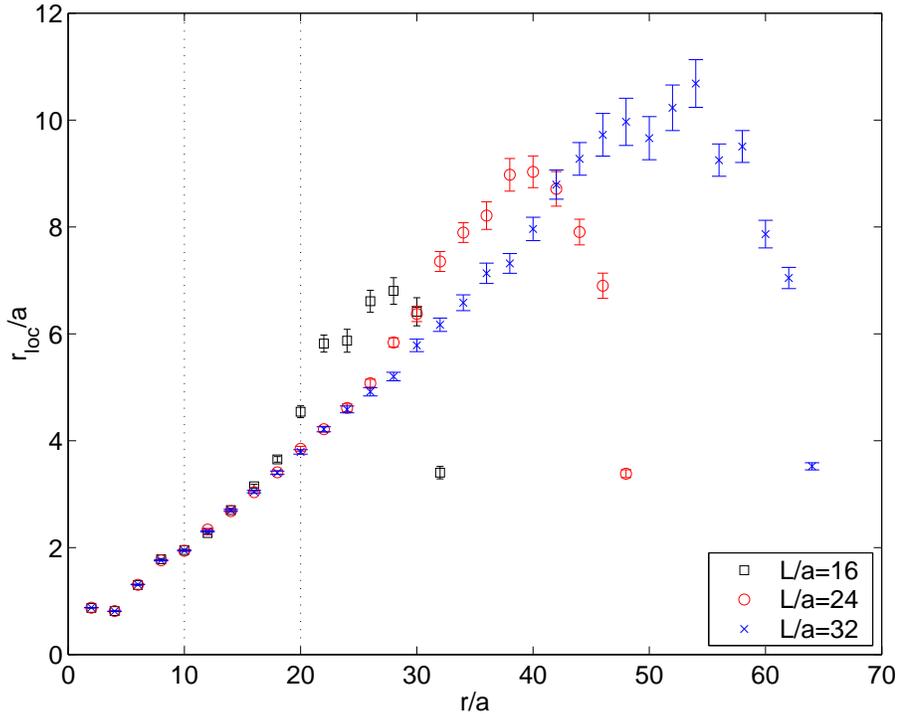}
 \end{center}
 \vspace{-0.5cm}
 \caption{Effective localization ranges $\rloc(r)$ at $\beta=6.0,\;am=0.01$
 for different volumes. The vertical dotted lines correspond to
 distances from the source smaller than the minimal distances
 $\;r_{\rm min}(L)\approx L\;$ 
 at which finite volume effects for $L/a=16,\;24$ become sizeable.}
 \label{f_rlocb60}
\end{figure}

In \fig{f_fr} we show in a semilogarithmic plot
the results for $\ev{f(r)}$ as function of the taxi driver
distance $r$ in units of $r_0$ at the three different
$\beta$ values for the volume $L\mG\approx6$.
For a better comparison of the curves we normalized $\ev{f(r)}$ by
the (linearly interpolated) value $\ev{f(r_{\rm ref})}$, where
the physical distance $r_{\rm ref}/r_0=3.72$ corresponds to the rightmost
vertical dotted line. \fig{f_fr} shows remarkable scaling as the continuum
limit is approached, which means that there is a physical scale of
``non-localization'' $\rloc$.

For small distances $r\ll\Lambda_{\rm QCD}^{-1}$ the form of $\ev{f(r)}$ is
dictated by perturbation theory and it is expected to follow the
polynomial (in $1/r$) result in the free theory \eq{freethcont}.
For large distances $r$ we assume
\bes
 \ev{f(r)} & \propto & {\rm e}^{-r/\rloc(r)} \,.
\ees
The inverse localization range $\rloc(r)^{-1}$ can be computed by taking
two consecutive values of $r$ to extract the exponential. The results for
three volumes at $\beta=6.0$ are shown in \fig{f_rlocb60}.
It is clear that
the decay of $\ev{f(r)}$ is not described by a single exponential.
On each of the volumes
$\rloc(r)$ has a ``bump'' for large taxi driver distances,
which becomes higher as the volume gets larger.
This is a finite volume effect and is discussed in Appendix \ref{finivolapp}.
\begin{figure}[t]
 \begin{center}
     \includegraphics[width=12cm]{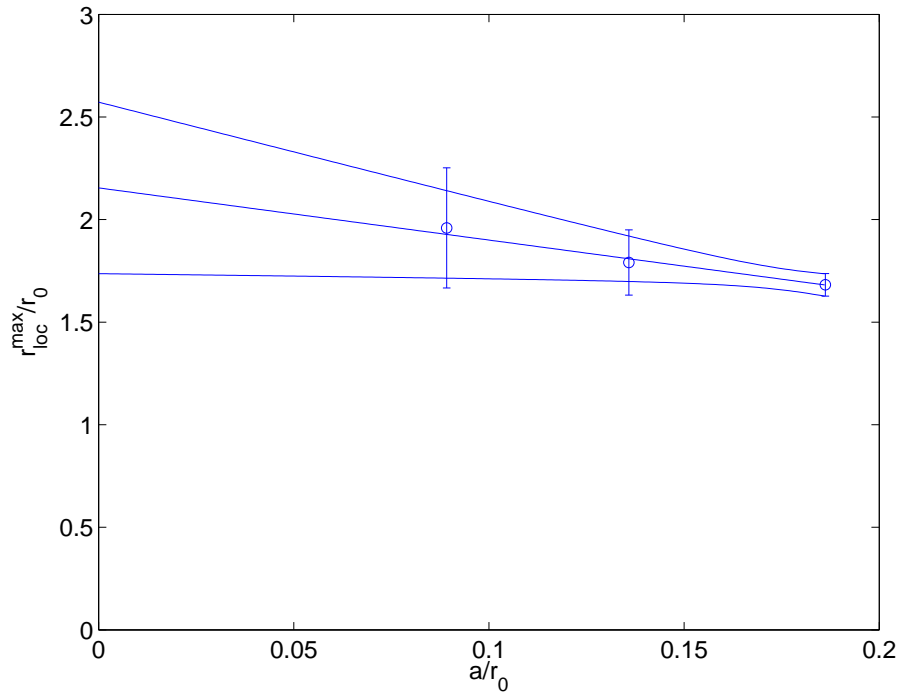}
 \end{center}
 \vspace{-0.5cm}
 \caption{The continuum limit of the upper bound on the localization range
 in the volume $L\mG\approx6$.}
 \label{f_uprloc}
\end{figure}

We adopt two strategies to define a localization range of the
operator $D$ in the continuum limit.
The first is to take for the
physical volume $L\mG\approx6$ the largest value of $\rloc(r)$,
which we denote by $\rlocm$, at each $\beta$ value.
Changing the volume will slightly change this value but the
volume $L\mG\approx6$ can be considered typical for present quenched
simulations. The results are shown in \fig{f_uprloc} and give a
continuum limit value
\bes
 \frac{\rlocm}{r_0} & = & 2.15(42) \quad \mbox{for}\quad L\mG\approx6 \,,
\ees
or $\rlocm\mG=2.8(6)$. Compared to the analytic bound given in \eq{rlocbound} the
values of $\rlocm$ that we obtain at the three $\beta$ values are
about 40 times smaller. The fact that the analytic bound is poorly saturated is in
agreement with similar observations made in Ref. \cite{Hernandez:1998et}.
\begin{figure}[ht]
 \begin{center}
     \includegraphics[width=12cm]{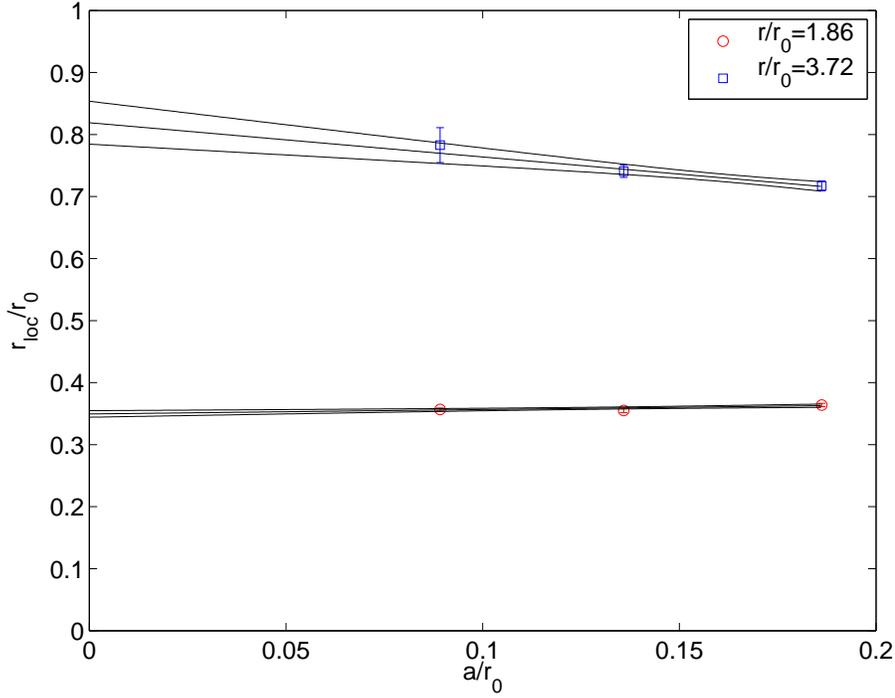}
 \end{center}
 \vspace{-0.5cm}
 \caption{The continuum limit of the localization range computed at
 two constant physical distances from the source.}
 \label{f_rlocfixr}
\end{figure}

The second strategy is to compute the localization range at a physical value
of the taxi driver distance from the source. We take two distances,
$r/r_0=1.86$ and twice this value $r/r_0=3.72$, which are marked in
\fig{f_fr}, \fig{f_rlocb60} and \fig{f_finivol} by the vertical dotted lines.
At $\beta=6.0$
there are no finite size effects in $\rloc(r)$ computed at these
two distances on lattices of size $L/a=24$ ($L\mG\approx6$),
as can be seen by comparing
to results on the larger volume $L/a=32$ in \fig{f_finivol}.
For $\rloc(r/r_0=1.86)$ we checked at the other $\beta$ values
that already in the volume $L\mG\approx4$ there are no finite size effects.
We take the lattice volume $L\mG\approx6$
and perform a linear interpolation of $\rloc(r)$ to get its values at the desired
distances for the three $\beta$ values.
We obtain the results shown
in \fig{f_rlocfixr} giving the continuum limit values
\bes
 \frac{\rloc(r/r_0=1.86)}{r_0} & = & 0.350(5) \,, \label{rlocfixr1} \\
 \frac{\rloc(r/r_0=3.72)}{r_0} & = & 0.819(35)\,, \label{rlocfixr2}
\ees
or $\rloc(r/r_0=1.86)\mG=0.455(12)$ and $\rloc(r/r_0=3.72)\mG=1.06(5)$.
We remind that a local operator would imply $\rloc(r)={\rm O}(a)$ for
all distances $r$.

\section{Conclusions}

We have studied the locality problem for a theory defined 
by taking the square root of the staggered fermion operator.
In terms of the fundamental fermion fields, this 
operator needs to be local 
in order to prove causality and unitarity of the so defined theory.
In our work, 
we proved that in the free field limit such a theory
is non-local at the scale of the inverse quark mass. Adding gauge fields, simulations in
the quenched approximation have revealed that the ``non-localization'' range
is of the order of the inverse Goldstone pion mass in the continuum limit.
We do not expect that taking the square root of staggered fermion operators
constructed with improved actions such as Asqtad \cite{Gottlieb:2003bt} or
HYP \cite{Hasenfratz:2001hp} will change qualitatively and even quantitatively
the situation, since smearing makes configurations even closer to the free case.

How can we interpret our results, also in the context of present day
simulations?
We have studied a candidate theory for two tastes of staggered fermions
obtained by taking the square root of the staggered fermion operator in
the action and the {\em corresponding} Green's functions. Our work
shows that such a theory is non-local in the continuum limit at the
scale of the Goldstone pion Compton wavelength, i.e. the lightest
particle in the theory. Hence this solution of the ``two taste problem''
leads to an unacceptable field theory that has to be dismissed. This leaves
the question open, of course, 
whether there exists a local operator $D$, which provides a
Boltzmann weight equal to the square root of the staggered fermion
determinant.

Present dynamical staggered fermion simulations  
do {\em not} use
this setup since only the square root of the {\em determinant} is employed. 
The corresponding action  
$S=1/2\tr\ln\mdme$ is simulated with the R algorithm
\cite{Gottlieb:1987mq} with the zero step-size limit or
$S=-\phid(\mdme)^{-1/2}\phi$ 
in terms of a {\em pseudo}fermion field $\phi$ with an exact algorithm
\cite{Aoki:2002xi,Clark:2003na}. We emphasize that these simulations 
generate the correct Boltzmann weight since the non-local bosonic formulation is 
only a technical trick to perform the numerical simulations. 
If a local operator $D$ reproducing the square root of the staggered fermion
determinant is found then the configurations generated by present algorithms
are safe. 

There still remains, however, the problem of unitarity. 
The point is a mismatch between the operators used for valence and sea quarks.
The valence quarks are discretized through the four taste staggered operator,
while the sea quarks would be introduced through the ``still to be found''
two taste local operator $D$. We believe that at this point the recovery of the
optical theorem is an open problem, which needs a clarification.
The operator $D$, acting on the fundamental fermion fields, 
should dictate which are the appropriate {\em Green's functions} for
a two taste theory. The Green's functions that are used at present
are most likely not the ones that correspond to 
the local operator $D$.
\\
{\bf Acknowledgements:} We thank A. Alexandru, T. DeGrand,
A. Hasenfratz, P. Hasenfratz, R. Hoffmann, A. Kronfeld, M. L{\"uscher},
F. Niedermayer and U. Wolff for precious discussions and suggestions.
The simulations in the present work have been carried out using
the MILC collaboration's public lattice gauge theory code. We also
thank the computer center at DESY Zeuthen and Humboldt University for their
professional assistance.

\begin{appendix}
\section{A polynomial expansion of the square root \label{polyexp}}

\newcommand{\Om}{\Omega}

Consider the generating function of the Gegenbauer polynomials
$C_n^\gamma(z)$
\bes
      (1 + t^2 - 2tz)^{-\gamma} &=& \sum_{n=0}^\infty t^n C_n^\gamma(z)
                        \label{Cn-genfct}
\ees
They are defined for any (fixed) $\gamma \in \mathbb{C}$. Their natural
domain is $z \in [-1,1]$. 
Many details about Gegenbauer polynomials can be found in \cite{HTFII},
section 10.9. We only list the recursion and a bound:
\bes
      (n+1) C_{n+1}^\gamma(z) &+& (n+2\gamma-1) C_{n-1}^\gamma(z)
            \;=\; 2(n+\gamma) \, z C_n^\gamma(z)      \label{Cn-recur}\\
      |C_n^\gamma(z)| &\le& \left( n+2\gamma-1 \atop n \right)
                  \qquad \gamma > 0                   \label{Cn-bound}
\ees

The power series (\ref{Cn-genfct}) converges for $|t| < 1$.
This may be used as a starting point for
{\em fractional inversion}\cite{Bunk:1998wj}. Here we specialize to the case
of $\gamma = -1/2$, which provides an expansion of the {\em square root}:
\bes
      \sqrt{1 + t^2 - 2tz} &=& \sum_{n=0}^\infty t^n S_n(z) \label{root-ex}\\
      S_n(z) &\equiv& C_n^{-1/2}(z) \nn
\ees
The first few polynomials $S_n(z)$ are
\bes
      S_0(z) &=& 1                  \nn\\
      S_1(z) &=& -z                 \nn\\
      S_2(z) &=& \frac{1}{2} (1 - z^2)    \nn
\ees
We find $S_n(\pm 1) = 0$ for $n \ge 2$. In fact the recursion (\ref{Cn-recur})
allows us to prove the useful relation
\bes
      S_n(z) &=& \frac{1 - z^2}{n(n-1)} \; C_{n-2}^{3/2}(z)
                  \qquad n \ge 2    \label{Sn-Cn}
\ees
As an immediate consequence, the bound (\ref{Cn-bound}) provides the estimate
\bes
      |S_n(z)| &\le& \frac{1}{2} \qquad n \ge 2     \label{Sn-bound}
\ees

In applications of the expansion (\ref{root-ex}), there is particular
interest in the {\em remainder} $\Om_n(z)$ of the series truncated after
the term $\sim t^n$
\bes
      \Om_n(z) &=& \sum_{k=n+1}^\infty t^k S_k(z)     \label{Om-def}
\ees
A simple (uniform) estimate follows immediately with the aid of
(\ref{Sn-bound}):
\bes
      |\Om_n(z)| &\le& \frac{1}{2} \sum_{k=n+1}^\infty |t|^k
            = \frac{1}{2} \frac{|t|^{n+1}}{1 - |t|}
                        \qquad n \ge 1              \label{Om-est1}
\ees
The coefficient of $|t|^{n+1}$ diverges as $|t| \to 1$.

We will prove the stronger (and more realistic) bound
\bes
      |\Om_n(z)| &\le& \frac{2}{n\pi} \; |t|^{n+1}
                        \qquad n \ge 1          \label{Om-est2}
\ees
for $z \in [-1,1]$ and $t \in (-1,1)$.

Assume $t \in (0,1)$ and let $z = \cos\phi$. The relation 
(\ref{Sn-Cn}) translates the integral representation of $C_n^{3/2}$
(\cite{HTFII}, section 10.9, eq.(31)) into
\bes
      S_n(\cos\phi) &=& \sin^2\phi \, 
                  \int_0^\pi \frac{d\varphi}{\pi} \, \sin^2\varphi \,
                  (\cos\phi + i \sin\phi \cos\varphi)^{n-2}
                        \qquad n \ge 2          \nn
\ees
This allows us to perform the summation in $\Om_n(z)$ eq.(\ref{Om-def})
(if $n \ge 1$)
\bes
      \Om_n(\cos\phi)
      &=& \sin^2\phi \; t^{n+1}
            \int_0^\pi \frac{d\varphi}{\pi} \, \sin^2\varphi \,
            \frac{(\cos\phi + i\sin\phi \cos\varphi)^{n-1}}
                 {1 - t(\cos\phi + i\sin\phi \cos\varphi)}        \nn
\ees
At the end points of the integration, the denominator takes values
$1 - t e^{\pm i\phi} \equiv \rho e^{\mp i\delta}$, i.e. we
reparametrize $(t,\phi) \to (\rho,\delta)$, with $\rho > 0$ and
$\delta \in [0, \pi/2)$
\bes
      \rho &=& |1 - t e^{i\phi}| = \sqrt{1 + t^2 - 2t\cos\phi}
                                                \nn\\
      \cos\delta &=& (1 - t\cos\phi) / \rho     \label{rho-delta} \\
      \sin\delta &=&  (t \sin\phi) / \rho       \nn\\
 \Rightarrow\quad
      \Om_n(\cos\phi) &=& t^{n-1} \, \rho \sin^2\delta
            \int_0^\pi \frac{d\varphi}{\pi} \, \sin^2\varphi \,
            \frac{(\cos\phi + i\sin\phi \cos\varphi)^{n-1}}
                 {\cos\delta - i\sin\delta \cos\varphi}     \nn
\ees
Rewrite this as a contour integral over $u$:
\bes
      u &\equiv& \cos\delta - i\sin\delta \cos\varphi             \nn\\
 \Rightarrow\quad
      1 - \rho u &=& t(\cos\phi + i\sin\phi \cos\varphi)          \nn\\
 \Rightarrow\quad
      \Om_n(\cos\phi) &=& \rho \int_{e^{-i\delta}}^{e^{i\delta}}
            \frac{du}{\pi i u} \, \sqrt{1 + u^2 - 2u\cos\delta} \;
            (1 - \rho u)^{n-1}                                    \nn\\
      &=& \rho \int_{e^{-i\delta}}^{e^{i\delta}} \frac{du}{\pi i u} \,
            Q(u) \; (1 - \rho u)^{n-1}                            \nn\\
 \mbox{with}\quad
      Q(u) &\equiv& \sqrt{1 + u^2 - 2u\cos\delta}                 \nn\\
      &=& \sqrt{(u-e^{i\delta})(u-e^{-i\delta})}                  \nn
\ees
The cut of the square root is chosen on the negative real axis.
$Q(u)$ has branch points at $u = e^{\pm i\delta}$, with vertical cuts
going out to $\pm i\infty$. The contour of integration connects the
two branch points across the real axis. \\
Integrate by parts, using $Q(u) = 0$ at the end points:
\bes
      \Om_n(\cos\phi) &=& \frac{1}{n}
            \int_{e^{-i\delta}}^{e^{i\delta}} \frac{du}{\pi i} \,
            \left( \frac{d}{du} \frac{Q(u)}{u} \right) \; (1 - \rho u)^n
\ees
This integral is finally evaluated along the circle
\bes
      u(\varphi) &=& \frac{1 + e^{2i\varphi}}{2 \cos\delta}
      \;=\; \frac{\cos\varphi}{\cos\delta} \, e^{i\varphi}
            \qquad \varphi \in [-\delta, \delta]            \nn\\
 \Rightarrow\quad
      Q(u) &=& \frac{e^{i\varphi}}{\cos\delta}
            \sqrt{\cos^2\varphi - \cos^2\delta}             \nn\\
 \Rightarrow\quad
      \frac{Q(u)}{u} &=& \frac{\sqrt{\cos^2\varphi - \cos^2\delta}}
            {\cos\varphi}                                   \nn\\
 \Rightarrow\quad
      \Om_n(\cos\phi) &=& \frac{1}{n \pi i}
            \int_{-\delta}^\delta d\varphi \,
            \left( \frac{d}{d\varphi} \,
            \frac{\sqrt{\cos^2\varphi - \cos^2\delta}}{\cos\varphi}
            \right) [1 - \rho u(\varphi)]^n                 \nn
\ees
Along the contour of integration, 
\bes
      |1 - \rho u(\varphi)| &\le& t       \nn
\ees
Furthermore, $Q(u)/u$ is an even function of $\varphi$, which vanishes
at the end points and reaches a maximum of $\sin\delta$ in the middle.
In this way we estimate
\bes
      |\Om_n(\cos\phi)| &\le& \frac{t^n}{n \pi}
            \int_{-\delta}^\delta d\varphi \,
            \left| \frac{d}{d\varphi} \,
            \frac{\sqrt{\cos^2\varphi - \cos^2\delta}}{\cos\varphi}
            \right|                             \nn\\
      &\le& \frac{2 t^n}{n\pi} \, \sin\delta    \nn
\ees
Inspection of (\ref{rho-delta}) shows that $\sin\delta \le t$,
which proves (\ref{Om-est2}) for $t \in (0,1)$.

The case of $t \in (-1,0)$ follows due to the invariance of $\Om_n(z)$
w.r.t. $t \to -t, z \to -z$.

\renewcommand{\Re}{\mbox{Re}\,}

\section{The square root in the free case \label{sqrtfree}}

\subsection{Lattice computation}

On an infinite lattice in $d$ dimensions (lattice spacing $a$), we
study functions of the operator (matrix)
\be
       m^2 - \Delta
\ee
which reads in momentum space:
\bes
      \left( m^2 - \Delta \right) (p)
      &=& m^2 + \sum_\mu 2 a^{-2} (1 - \cos a p_\mu)              \nn\\
      &=& m^2 + \sum_\mu \frac{4}{a^2} \sin^2 \frac{a p_\mu}{2}   \nn\\
      -\pi/a \le p_\mu \le \pi/a                                  \nn
\ees
Start from the "generalized propagator" with $s \in \mathbb{C}$:
\bes
      \left( m^2 - \Delta \right)^{-s}(x)                         \nn
      &=& a^{2s} \, \int_{-\pi/a}^{\pi/a} \frac{d^d p}{(2\pi)^d} \;
            \frac{e^{ipx}}{[a^2 m^2 + \sum_\mu 2 (1 - \cos a p_\mu)]^s} \nn
\ees
Use the standard trick (valid for $A > 0$ and $\Re s > 0$)
\bes
      A^{-s} &=& \frac{1}{\Gamma(s)} \int_0^\infty dt \, t^{s-1} e^{-tA} \nn
\ees
to factorize the $p$ integrations. They lead to {\em modified Bessel functions
of the first kind} $I_n$:
\bes
      \int_{-\pi/a}^{\pi/a} \frac{dp_\mu}{2\pi}
            \exp\{ip_\mu x_\mu + 2t\cos a p_\mu\}     \nn
      &=& a^{-1} I_{x_\mu/a} (2t)   \nn
\ees
\bes
  \Rightarrow\quad
      \left( m^2 - \Delta \right)^{-s}(x)
      &=& \frac{a^{2s-d}}{\Gamma(s)} \int_0^\infty dt \,
            t^{s-1} e^{-t(a^2 m^2 + 2d)} \prod_\mu I_{x_\mu/a} (2t)
                                                \label{eq-gplatt}
\ees
For $t \to \infty$, $I_n(2t) \sim t^{-1/2} e^{2t}$ and the integral converges
if $m \ne 0$. In case of $m = 0$, we need $\Re s < d/2$.

For $t \to 0$: $I_n(2t) \sim t^{|n|}$. In case of $x \ne 0$, the product of
Bessel functions vanishes at least linearly in $t$. As a consequence,
eq.(\ref{eq-gplatt}) is valid for $\Re s > -1$ and can be evaluated at
$s = -1/2$:
\bes
      \left( m^2 - \Delta \right)^{1/2}(x)            \nn
      &=& \frac{a^{-d-1}}{\Gamma(-1/2)} \int_0^\infty dt \,
            t^{-3/2} e^{-t(a^2 m^2 + 2d)} \prod_\mu I_{x_\mu/a} (2t)    \\
      && (x \ne 0)            \label{eq-rootx}
\ees
(Note: $\Gamma(-1/2) = -2\sqrt{\pi}$).

For the special case of $x = 0$, continuation to $s = -1/2$ requires to
subtract a term in the integrand:
\bes
      I_0(2t)^d &=& \left[ I_0(2t)^d - 1 \right] + 1  \nn\\
      \int_0^\infty dt \, t^{s-1} e^{-t(a^2 m^2 + 2d)}
      &=& \Gamma(s) (a^2 m^2 + 2d)^{-s}               \nn\\
  \Rightarrow\quad
      \left( m^2 - \Delta \right)^{1/2}(0)
      &=& a^{-d-1} (a^2 m^2 + 2d)^{1/2}                \label{eq-rootx0} \\
      && + \frac{a^{-d-1}}{\Gamma(-1/2)} \int_0^\infty dt \,
            t^{-3/2} e^{-t(a^2 m^2 + 2d)} \left[ I_0(2t)^d - 1 \right]  \nn
\ees
\begin{figure}[t]
 \begin{center}
     \includegraphics[width=10cm]{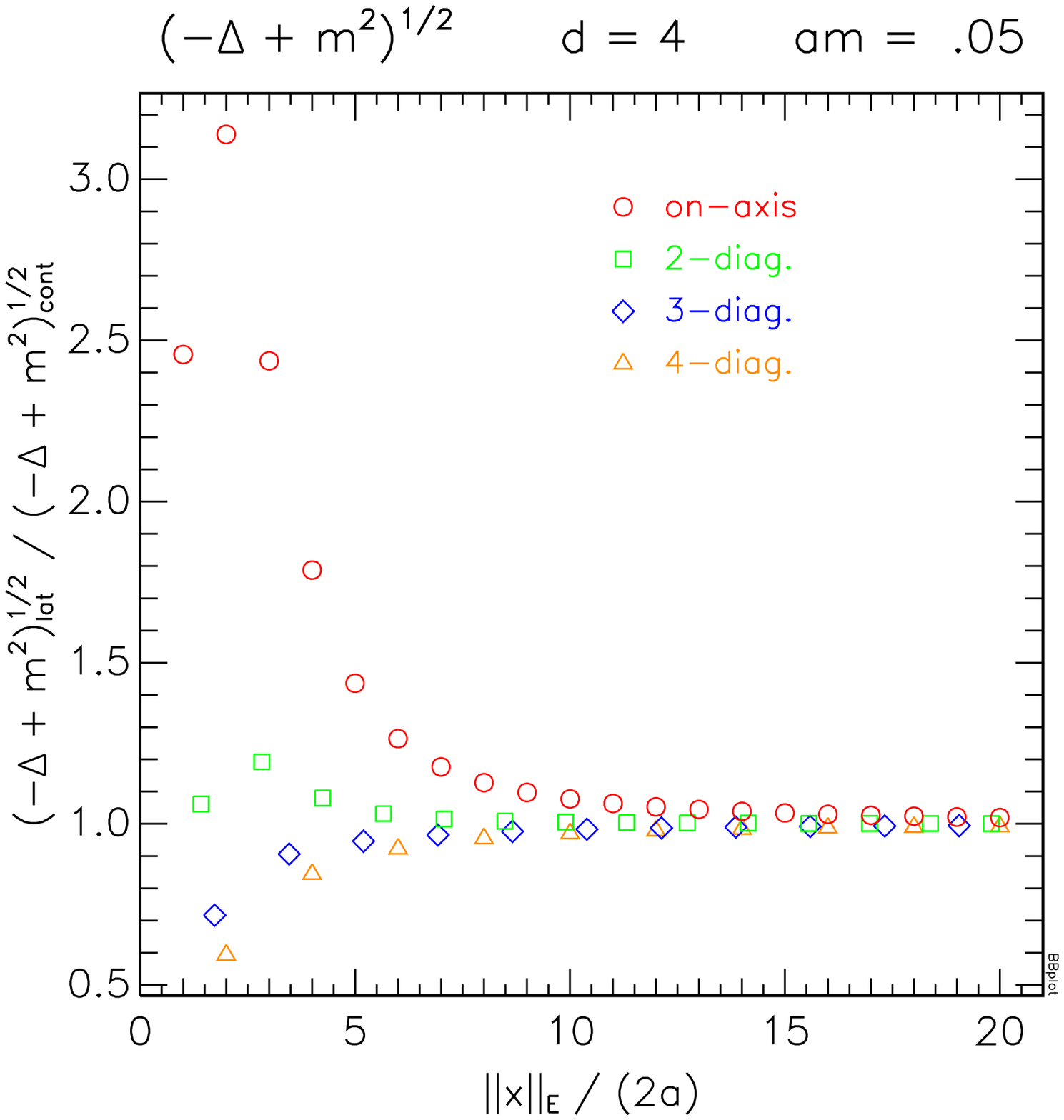}
 \end{center}
 \vspace{-0.5cm}
 \caption{Comparison of the lattice \eq{eq-rootx} and continuum \eq{eq-rootcont}
 results.\label{f_latroot}}
\end{figure}

\subsection{Continuum calculation}

Compute the "generalized propagator" again:
\bes
      \left( m^2 - \Delta \right)^{-s} (x)
      &=& \int_{-\infty}^\infty \frac{d^d p}{(2\pi)^d}
            \frac{e^{ipx}}{(m^2 + p^2)^s}                               \nn\\
      &=& \frac{1}{\Gamma(s)} \int_0^\infty dt \, t^{s-1}
            \int \frac{d^d p}{(2\pi)^d} \, e^{ipx - (m^2 + p^2)t}       \nn\\
      &=& \frac{(4\pi)^{-d/2}}{\Gamma(s)} \int_0^\infty dt \, t^{s-1-d/2}
            e^{-m^2 t - x^2/(4t)}                                       \nn\\
\ees
This time, we end up with modified Bessel functions of the {\em second}
kind $K_\nu$:
\bes
      \int_0^\infty dt \, t^{s-1-d/2} e^{-m^2 t - x^2/(4t)}
      &=& \left( \frac{||x||}{2m} \right)^{s-d/2} \int_0^\infty du \,
            u^{s-1-d/2} e^{-m||x|| (u + u^{-1})/2}                      \nn\\
      &=& \left( \frac{||x||}{2m} \right)^{s-d/2} 2 K_{d/2-s}(m||x||)   \nn\\
  \Rightarrow\quad
      \left( m^2 - \Delta \right)^{-s} (x)
      &=& \frac{(4\pi)^{-d/2}}{\Gamma(s)}
            \left( \frac{||x||}{2m} \right)^{s-d/2} 2 K_{d/2-s}(m||x||)
\ees
The derivation is valid for $x \ne 0$ and $0 < \Re s < d/2$, but analytic
continuation to $s = -1/2$ is obvious:
\bes
      \left( m^2 - \Delta \right)^{1/2} (x)
      &=& \frac{(4\pi)^{-d/2}}{\Gamma(-1/2)}
            \left( \frac{||x||}{2m} \right)^{-(d+1)/2} 2 K_{(d+1)/2}(m||x||) \\
      &&          (x \ne 0)   \nn
\ees
In {\em four} dimensions, we make use of the fact that Bessel functions with
half--integral index are elementary, e.g.
\bes
      K_{5/2}(z) &=& \left( \frac{\pi}{2z} \right)^{1/2} e^{-z}
            \left( 1 + \frac{3}{z} + \frac{3}{z^2} \right)              \nn\\
  \Rightarrow\quad
      \left( m^2 - \Delta \right)^{1/2} (x)
      &=& - \frac{1}{4\pi^2} \, \frac{m^2}{||x||^3} \, e^{-m||x||} \,
            \left( 1 + \frac{3}{m||x||} + \frac{3}{m^2 ||x||^2} \right) \nn\\
      &=& - \frac{3}{4\pi^2} \, ||x||^{-5} \, e^{-m||x||} \,
            \left( 1 + m||x|| + m^2 ||x||^2/3 \right)                   \nn\\
      &&          (d = 4)                 \label{eq-rootcont}
\ees

\subsection{Comparison}

The result in \eq{eq-rootcont} is the (formal) continuum limit of the lattice
expression \eq{eq-rootx}. We expect
\bes
      \left( m^2 - \Delta \right)_{lat}^{1/2}(x)
      &\approx& \left( m^2 - \Delta \right)_{cont}^{1/2}(x)  \label{compa}\nn
\ees
In fact, it has been verified numerically that the lattice expression
approaches the continuum result with increasing $||x||/a$, using the
{\em Euclidean} distance $||x||_E$ on the lattice as well.
This is shown in \fig{f_latroot} for $am = 0.05$.

\section{Finite volume effects \label{finivolapp}}

When applying the staggered operator $\mdm$ hops to neighbors
are suppressed by a factor $1/(4(am)^2+8)$ with respect to the
static mass term.
So $aD\xi_c(x)$ in \eq{applyD} receives smaller contributions from
paths with a larger number of hops connecting $x$ with the source at $y$.
As the taxi driver distance $\norm{x-y}_1$ \eq{dtaxi} grows,
the relative weight
of path wrapping around the lattice grows. This growth though is expected to
be smaller in comparison on larger lattice sizes $L^{\prime}>L$
simply because the path ``around the world'' is
longer.
\begin{figure}[ht]
 \begin{center}
     \includegraphics[width=10cm]{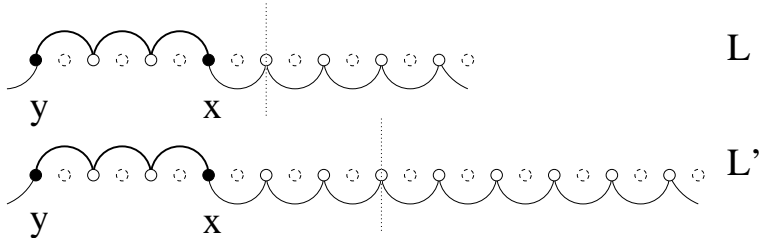}
 \end{center}
 \vspace{-0.5cm}
 \caption{Representation in one dimension of a finite volume effect. 
 The thickness of the lines representing the hops is proportional
 to the weight of their contribution in \eq{applyD}. The vertical
 dotted lines mark the largest possible taxi driver distance from
 the source at $y$. \label{f_hop}}
\end{figure}

\begin{figure}[t]
 \begin{center}
     \includegraphics[width=10cm]{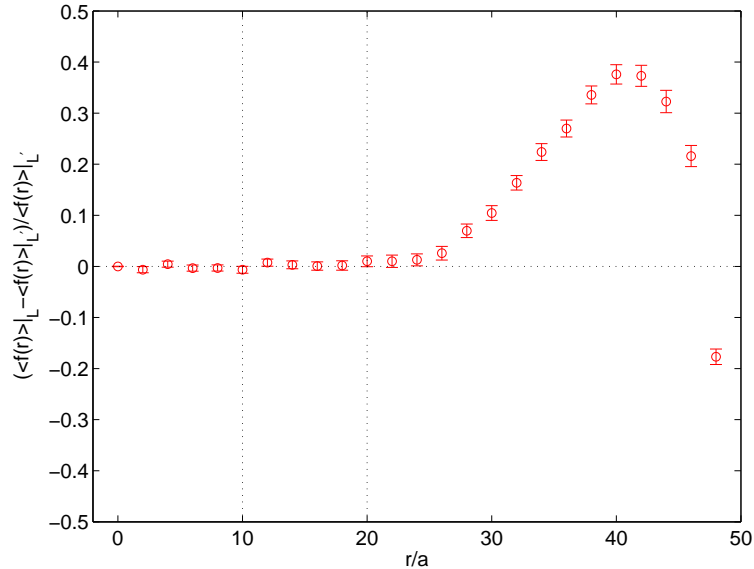}
 \end{center}
 \vspace{-0.5cm}
 \caption{Finite volume effect at $\beta=6.0\,,\;am=0.01$. Comparison
 of $\ev{f(r)}$ computed on $L/a=24$ with $L^{\prime}/a=32$ lattices
 for $0\le r\le2L$.
 \label{f_finivol}}
\end{figure}

\noindent
The situation is schematically represented in \fig{f_hop}.
Thus we expect
\bes
 \ev{f(r)}|_L \ge \ev{f(r)}|_{L^{\prime}} \,,\quad L^{\prime}>L \,, \quad
                                                 r>\mbox{some}\;r_{\rm min}
 \,. \label{finvoleff1}
\ees
This expectation is confirmed by the numerical results shown in 
\fig{f_finivol}, which are obtained at
$\beta=6.0\,,\;am=0.01$ comparing $\ev{f(r)}$ computed on $L/a=24$ with
$L^{\prime}/a=32$ lattices for $0\le r\le2L$.
We observe that the minimal distance at
which the finite size effects become sizeable is $r_{\rm min}\approx L$.
The vertical dotted lines in \fig{f_finivol} correspond to the distances
used in \eq{rlocfixr1} and \eq{rlocfixr2}. At these two values of $r$
finite volume effects are absent on the $L/a=24$ lattice.

The effect leading to \eq{finvoleff1} is the dominant (but not the only
one) finite volume effect. It is responsible for the ``bumps'' of the
effective localization range at large distances $r$
visible in \fig{f_rlocb60}, which vanish as the volume increases.
The size of this finite volume effect decreases as the mass $am$
increases.

Concluding this section we remark that the maximization operation in the definition
of $f(r)$ \eq{fr} solves the ambiguity for the case when points at the same taxi
driver distance but different Euclidean distances from the source are considered.
The maximum is presumably attained at the point with the smallest Euclidean
distance,
which is also the one least affected by the four-dimensional generalization of the
finite size effect depicted here in \fig{f_hop}.

\end{appendix}

\bibliography{ks}           %or whatever your .bib file is

\begin{thebibliography}{10}

\bibitem{Wilson:1975id}
K.G. Wilson,
\newblock New Phenomena In Subnuclear Physics. Part A. Proceedings of the First
  Half of the 1975 International School of Subnuclear Physics, Erice, Sicily,
  July 11 - August 1, 1975, ed. A.~Zichichi, Plenum Press, New York, 1977,
  p.~69, CLNS-321.

\bibitem{Kogut:1975ag}
J.B. Kogut and L. Susskind,
\newblock Phys. Rev. D11 (1975) 395.

\bibitem{Susskind:1977jm}
L. Susskind,
\newblock Phys. Rev. D16 (1977) 3031.

\bibitem{Rossi:1984kr}
P. Rossi, U. Wolff and D. Zwanziger,
\newblock Phys. Rev. D30 (1984) 2233.

\bibitem{Kluberg-Stern:1981wz}
H. Kluberg-Stern, A. Morel, O. Napoly and B. Petersson,
\newblock Nucl. Phys. B190 (1981) 504.

\bibitem{Kawamoto:1981hw}
N. Kawamoto and J. Smit,
\newblock Nucl. Phys. B192 (1981) 100.

\bibitem{Kluberg-Stern:1983dg}
H. Kluberg-Stern, A. Morel, O. Napoly and B. Petersson,
\newblock Nucl. Phys. B220 (1983) 447.

\bibitem{Golterman:1984cy}
M.F.L. Golterman and J. Smit,
\newblock Nucl. Phys. B245 (1984) 61.

\bibitem{Kilcup:1987dg}
G.W. Kilcup and S.R. Sharpe,
\newblock Nucl. Phys. B283 (1987) 493.

\bibitem{Neuberger:1998fp}
H. Neuberger,
\newblock Phys. Lett. B417 (1998) 141, hep-lat/9707022.

\bibitem{Jansen:2003nt}
K. Jansen,
\newblock (2003), hep-lat/0311039.

\bibitem{Toussaint:2001zc}
D. Toussaint,
\newblock Nucl. Phys. Proc. Suppl. 106 (2002) 111, hep-lat/0110010.

\bibitem{Bernard:2001av}
C.W. Bernard et~al.,
\newblock Phys. Rev. D64 (2001) 054506, hep-lat/0104002.

\bibitem{Gattringer:2003qx}
BGR, C. Gattringer et~al.,
\newblock Nucl. Phys. B677 (2004) 3, hep-lat/0307013.

\bibitem{Aoki:2002fd}
CP-PACS, S. Aoki et~al.,
\newblock Phys. Rev. D67 (2003) 034503, hep-lat/0206009.

\bibitem{Gottlieb:1987mq}
S. Gottlieb, W. Liu, D. Toussaint, R.L. Renken and R.L. Sugar,
\newblock Phys. Rev. D35 (1987) 2531.

\bibitem{Sharatchandra:1981si}
H.S. Sharatchandra, H.J. Thun and P. Weisz,
\newblock Nucl. Phys. B192 (1981) 205.

\bibitem{vandenDoel:1983mf}
C. van~den Doel and J. Smit,
\newblock Nucl. Phys. B228 (1983) 122.

\bibitem{Marinari:1981qf}
E. Marinari, G. Parisi and C. Rebbi,
\newblock Nucl. Phys. B190 (1981) 734.

\bibitem{Davies:2003ik}
HPQCD, C.T.H. Davies et~al.,
\newblock Phys. Rev. Lett. 92 (2004) 022001, hep-lat/0304004.

\bibitem{DeGrand:2003xu}
T. DeGrand,
\newblock (2003), hep-ph/0312241.

\bibitem{Neuberger:2004ft}
H. Neuberger,
\newblock (2004), hep-ph/0402148.

\bibitem{Bernard:1994sv}
C.W. Bernard and M.F.L. Golterman,
\newblock Phys. Rev. D49 (1994) 486, hep-lat/9306005.

\bibitem{Hernandez:1998et}
P. Hernandez, K. Jansen and M. L{\"u}scher,
\newblock Nucl. Phys. B552 (1999) 363, hep-lat/9808010.

\bibitem{Gottlieb:2003bt}
S. Gottlieb,
\newblock (2003), hep-lat/0310041.

\bibitem{Aoki:2002xi}
JLQCD, S. Aoki et~al.,
\newblock Comput. Phys. Commun. 155 (2003) 183, hep-lat/0208058.

\bibitem{Frezzotti:1997ym}
R. Frezzotti and K. Jansen,
\newblock Phys. Lett. B402 (1997) 328, hep-lat/9702016.

\bibitem{Clark:2003na}
M.A. Clark and A.D. Kennedy,
\newblock (2003), hep-lat/0309084.

\bibitem{Niedermayer:1998bi}
F. Niedermayer,
\newblock Nucl. Phys. Proc. Suppl. 73 (1999) 105, hep-lat/9810026.

\bibitem{Martin:1985yn}
O. Martin and S.W. Otto,
\newblock Phys. Rev. D31 (1985) 435.

\bibitem{Numerical-Recipes}
W.H. Press, B.P. Flannery, S.A. Teukolsky and W.T. Vetterling,
\newblock Numerical Recipes: The Art of Scientific Computing, 2nd ed.
  (Cambridge University Press, Cambridge (UK) and New York, 1992).

\bibitem{Matrix-Analysis}
R.A. Horn and C.R. Johnson,
\newblock Matrix Analysis (Cambridge University Press, Cambridge (UK) and New
  York, 1985).

\bibitem{Kalkreuter:1995ax}
T. Kalkreuter,
\newblock Phys. Rev. D51 (1995) 1305, hep-lat/9408013.

\bibitem{Kim:1995tg}
S. Kim and D.K. Sinclair,
\newblock Phys. Rev. D52 (1995) 2614, hep-lat/9502004.

\bibitem{Gupta:1991mr}
R. Gupta, G. Guralnik, G.W. Kilcup and S.R. Sharpe,
\newblock Phys. Rev. D43 (1991) 2003.

\bibitem{Kim:1999ur}
S. Kim and S. Ohta,
\newblock Phys. Rev. D61 (2000) 074506, hep-lat/9912001.

\bibitem{Sommer:1994ce}
R. Sommer,
\newblock Nucl. Phys. B411 (1994) 839, hep-lat/9310022.

\bibitem{Guagnelli:1998ud}
ALPHA, M. Guagnelli, R. Sommer and H. Wittig,
\newblock Nucl. Phys. B535 (1998) 389, hep-lat/9806005.

\bibitem{Wolff:2003sm}
ALPHA, U. Wolff,
\newblock Comput. Phys. Commun. 156 (2004) 143, hep-lat/0306017.

\bibitem{Hasenfratz:2001hp}
A. Hasenfratz and F. Knechtli,
\newblock Phys. Rev. D64 (2001) 034504, hep-lat/0103029.

\bibitem{HTFII}
A. Erd\'elyi ed.,
\newblock Higher Transcendental Functions, Vol. II (Robert E. Krieger
  Publishing Company, Malabar, Florida, 1981).

\bibitem{Bunk:1998wj}
B. Bunk,
\newblock Nucl. Phys. Proc. Suppl. B63 (1998) 952, hep-lat/9805030.

\end{thebibliography}
\bibliographystyle{h-elsevier}   %if you use h-elsevier.bst

\end{document}